\documentclass[aps,pra,twocolumn,showpacs]{revtex4}

\usepackage{graphicx}
\usepackage{subfigure}
\usepackage{bm}

\begin{document}

\title{Non-adiabatic loading of a Bose-Einstein condensate into the ground state of an optical lattice}

\author{A. S. Mellish}
 \email{amellish@physics.otago.ac.nz}
\author{G. Duffy}
\author{C. McKenzie}
\author{R. Geursen}
\author{A. C. Wilson}
\affiliation{%
Department of Physics, University of Otago, Dunedin, New Zealand
}

\date{\today}

\begin{abstract}
We present a scheme for rapidly loading a Bose-Einstein condensate into a single Bloch state of a weak optical lattice at a quasi-momentum of $q = \hbar k$. Rabi cycling of the Bose condensate between momentum states is modified by a phase shift applied to the lattice. By appropriately choosing the magnitude and timing of the phase shift, we demonstrate nearly perfect loading of the lattice ground state, the signature of which is an abrupt halt to the Rabi cycling.
\end{abstract}


\maketitle

Recent experiments involving the manipulation of Bose-Einstein condensates (BECs) \cite{Anderson1995, Davis1995, Pethick2002} in optical lattices highlight some of the remarkable features of quantum states of matter. The first experiment involving a Bose condensate and an optical lattice was performed by Anderson and Kasevich \cite{Anderson1998} observing the interference of atomic de Broglie waves tunneling through a lattice due to gravity. Optical lattices at the Bragg condition have been used to demonstrate a coherent beam splitter for de Broglie waves \cite{Kozuma1999} and to make spectroscopic measurements of condensate momentum distributions \cite{Stenger1999}. A diverse range of other experiments has been performed using BECs in optical lattices \cite{Ovchinnikov1999,Orzel2001,Hensinger2001,Cataliotti2001,Greiner2001,Cristiani2002,Greiner2002a,HeckerDenschlag2002}, involving a wide variety of parameters and techniques. One reason for this strong interest in optical lattices is their potential application in quantum information processing \cite{Jaksch1999, Hemmerich1999, Brennen1999}, where precise control and negligible decoherence are essential. For use in quantum computing the condensate must first be loaded into the ground state of the lattice. This is typically done by adiabatically increasing the lattice depth (see, for example, Refs.~\cite{Greiner2001, HeckerDenschlag2002}) or evaporative cooling in a hybrid trap \cite{Burger2001}. However, for the case where the quasi-momentum $q$ approaches the boundary of the first Brillouin zone (at $q = \pm \hbar k$, where $k=2\pi/\lambda$ is the lattice wave-vector), the adiabatic loading method fails due to the initial degeneracy of the Bloch bands.

In this paper we report a new scheme for rapid (non-adiabatic) loading into the ground state of an optical lattice at $q=\hbar k$.  A Bose condensate is placed in a weak moving lattice which induces Rabi cycling between two momentum states, 0 and 2$\hbar k$, via Bragg scattering \cite{Blakie2000}. We show that by simply ``jumping'' the phase of the periodic standing wave potential after a $3\pi/2$ pulse, we are able to transfer the entire population to the lattice ground state.  We present a convenient geometrical representation (using the Bloch sphere \cite{Allen1975}) which is a useful tool in predicting the outcome of our two-state manipulations.  Our atom-optical scheme is adapted from the methods developed by Hartman and Hahn \cite{Hartmann1962} to generate spin-locked states in nuclear magnetic resonance \cite{Abragam1961} and Bai {\it et al.}\ \cite{Bai1985} to produce stationary dressed atomic states in resonance fluorescence \cite{Cohen-Tannoudji1977}.  A similar adaptation of the NMR technique was recently used by Schmidt-Kaler {\it et al.}\ \cite{Schmidt-Kaler2003} to demonstrate a Cirac-Zoller controlled-NOT quantum gate with trapped ions \cite{Cirac1995}. 

The potential of a one-dimensional optical lattice formed by two counterpropagating laser beams is given by
\begin{equation}
 V(x,t) = \frac{V_0}{2}[1 + \cos(2kx + \delta t)]
\end{equation}
\noindent where $V_0$ is the lattice depth and $\delta$ is the frequency difference between the two beams. A Bose-Einstein condensate suddenly loaded into such a potential can be described in terms of Bloch eigenstates, which are a superposition of plane-wave states separated in momentum space by $2\hbar k$. When loading into a weak lattice at the Bragg condition $q=\hbar k$ the eigenstates can be approximated as
\begin{eqnarray} 
	\label{eq:phi0}
	\phi_{0} &=& \frac{1}{\sqrt 2}(e^{-ikx}+e^{ikx}), \\
	\label{eq:phi1}
	\phi_{1} &=& \frac{1}{\sqrt 2} (e^{-ikx} - e^{ikx})e^{i\omega t}
\end{eqnarray}
\noindent in the rest frame of the lattice. The standard two-photon Rabi frequency $\omega$ \cite{Metcalf1999} at the Bragg condition is proportional to the energy difference between the eigenstates. In the low-depth limit (which applies here) this energy gap is half the depth of the optical lattice \cite{Peik1997}.
Our approximation relies on the assumptions that the initial BEC is a plane wave, that only the lowest two eigenstates ($\phi_0$ and $\phi_1$) are involved, which is valid for a weak lattice, and that mean-field effects can be ignored. After loading at these conditions the system $\Psi$ is in an equal superposition of the two eigenstates \cite{rapidload} such that
\begin{equation} \label{eq:psi0}
	\Psi(t)  = \frac{1}{2}(1 + e^{i\omega t})e^{-ikx} + \frac{1}{2}(1 - e^{i\omega t})e^{ikx},
\end{equation}
\noindent leading to Rabi flopping between the two plane-wave states $e^{-ikx}$ and $e^{ikx}$ at frequency $\omega$.

If a sudden phase shift $\theta$, is applied to the lattice at $t=t_{\theta}$ then this is equivalent to shifting both plane wave states by $\theta/2$ (i.e.\ $kx \to kx + \theta/2)$ and keeping the origin of the lattice unchanged so the wavefunction becomes
\begin{equation} \label{eq:psiphase}
	\Psi(t=t_{\theta})  = \frac{1}{2}(1 + e^{i\omega t_{\theta}})e^{-ikx}e^{-i\theta/2} + \frac{1}{2}(1 - e^{i\omega t_{\theta}})e^{ikx}e^{i\theta/2}.
\end{equation}
\noindent Since we have chosen our reference frame so that the lattice remains stationary, equations (\ref{eq:phi0}) and (\ref{eq:phi1}) are still eigenstates of the system. To find the new population amplitudes we write
\begin{equation}
	\label{eq:psigen}
	\Psi(t=t_{\theta})  =  a_{0} \phi_{0} + a_{1} \phi_{1}
\end{equation}
\noindent and obtain
\begin{eqnarray}
	\label{eq:a0}
	a_{0} &=& \frac{1}{\sqrt 2}(\cos \frac{\theta}{2} - ie^{i\omega t_{\theta}}\sin \frac{\theta}{2}) \\
	\label{eq:a1}
	a_{1} &=& \frac{1}{\sqrt 2}(\cos \frac{\theta}{2} - ie^{-i\omega t_{\theta}}\sin \frac{\theta}{2}).
\end{eqnarray}

\noindent If we choose $t_{\theta}$ such that $\omega t_{\theta} = \pi/2$ $(3\pi/2)$ and choose $\theta = \pi/2$ $(-\pi/2)$, we get $a_{0} = 1$ and $a_{1} = 0$ so that the system is purely in the ground eigenstate. Or by selecting $\theta = -\pi/2$ $(\pi/2)$ for $\omega t_{\theta} = \pi/2$ $(3\pi/2)$ we get the entire population in the $\phi_1$ eigenstate. For other choices of $\omega t_{\theta}$ and $\theta$ the system remains in a superposition state.

In our experiment we begin by making a $^{87}$Rb Bose condensate containing approximately $2\times 10^{4}$ atoms in the $F=2$, $m_{F}=2$ hyperfine state. The condensate is formed (as described in Ref.~\cite{Martin1999}, but with minor modifications \cite{trapmod}) in a time-averaged orbiting potential trap with harmonic oscillation frequencies of $\omega_{r}/2\pi=71$~Hz radially and $\omega_{z}/2\pi=201$~Hz axially. After evaporation the trap is relaxed to $\omega_{r}/2\pi=32$~Hz and $\omega_{z}/2\pi=91$~Hz over a period of 200~ms. The condensate is released from the trap and an optical lattice is applied after 1.7~ms of free expansion, at which time the condensate extends across approximately 20 potential wells. Under these conditions we can ignore mean-field effects. The lattice is applied with pulse lengths between 100~$\mu$s and 350~$\mu$s. We probe the condensate using absorption imaging after an additional 8~ms (to allow the diffracted momentum components to separate).

The optical lattice is formed using a pair of counter-propagating laser beams derived from a single beam, which is detuned 4.49~GHz from the $5S_{1/2}, F=2 \to 5P_{3/2}, F'=3$ transition (giving a  spontaneous emission rate of 29~s$^{-1}$, i.e.\ negligible for our purposes). A double-pass acousto-optic modulator (AOM) is used in each beam for altering its frequency and switching the lattice on and off. The relative phase of the lattice beams is switched by altering the phase of the RF signal to one of the AOMs. The uncertainty in the phase shift is $\pm0.02$~rad. Each beam has a $1/e^{2}$ intensity halfwidth of 1.0~mm and an average intensity of 6.8~mW/cm$^{2}$. We set the frequency difference between the two lattice beams to give a quasi-momentum of $\hbar k$ (the first-order Bragg resonance condition \cite{Kozuma1999}). 

\begin{figure}
 \includegraphics[width=0.6\linewidth]{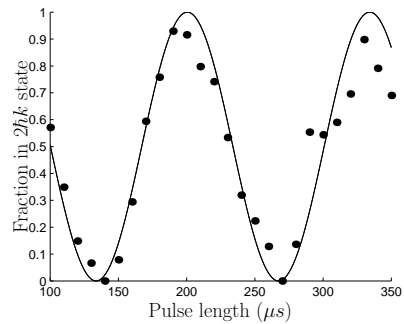}
 \caption{\label{fig:rabiplot} Resonant Rabi cycling of a Bose-Einstein condensate between 0 and $2\hbar k$ momentum states. The Rabi frequency $\omega/(2\pi)=7.5$ kHz gives a lattice depth of $\approx 4E_R$. Each point is an average of three data points and we estimate the uncertainty in the fraction to be $\pm$0.1. The solid line is a sinusoidal fit to the data.}
\end{figure}

Figure~\ref{fig:rabiplot} shows Rabi cycling between the 0 and $2\hbar k$ momentum states, as predicted by Eq.~\ref{eq:psi0}. We measure the fraction of the total number of atoms with $2\hbar k$ momentum for various lattice pulse lengths. The measured Rabi frequency is $2\pi\times (7.5\pm0.3$~kHz), from which we calculate the lattice depth to be approximately $4E_R$ (where $E_R=\hbar^{2} k^{2}/2m$ is the recoil energy). Based on a projection of the initial state onto the Bloch states, we predict more than $98\%$ population is in the lowest two eigenstates for this lattice depth, so our two-state approximation is valid.

\begin{figure}
  \subfigure[]{\includegraphics[width=0.6\linewidth]{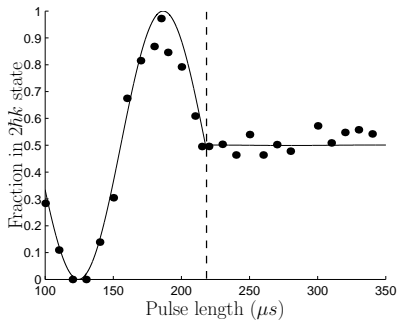}}
  \subfigure[]{\includegraphics[width=0.6\linewidth]{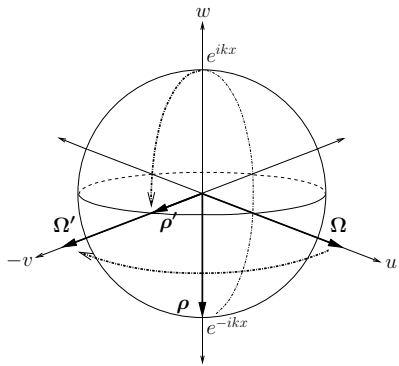}}
  \caption{\label{fig:rabiphase} (a) The effect of a $-\pi$/2 phase shift of the optical Bragg potential. The phase shift was applied at a time equivalent to a $3\pi$/2 pulse, indicated by the dashed line, resulting in loading of the ground eigenstate. The solid line is a  theoretical prediction. (b) A representation of the experiment on a Bloch sphere, where {\boldmath $\rho$} and  {\boldmath $\Omega$} are the initial Bloch and torque vectors respectively. Immediately after the phase shift they are represented by {\boldmath $\rho'$} and {\boldmath $\Omega'$}.}
\end{figure}

We now investigate the effect on the Rabi cycling of ``jumping'' the phase of the lattice. In the first case we choose the time for the phase jump $t_{\theta}$ such that $\omega t_{\theta}$ is equivalent to $3\pi/2$ (i.e.\ equal populations in the 0 and 2$\hbar k$ states) and perform a $-\pi/2$ phase shift. The results of this are shown in Fig.~\ref{fig:rabiphase}. The suppression of Rabi cycling after the phase shift demonstrates that we have successfully loaded the BEC into a single lattice eigenstate, the ground eigenstate, as predicted by Eqs.~(\ref{eq:a0}) and (\ref{eq:a1}). 

Since we have a two-level system, the Bloch sphere \cite{Allen1975} can be used to quickly predict the effect of any phase shift applied at any time in the Rabi cycle. We choose a Bloch sphere with the two plane-wave states at the poles because we measure the condensate momentum states (corresponding to $e^{-ikx}$ and $e^{ikx}$ in the lattice frame). The torque vector {\boldmath $\Omega$} \cite{Metcalf1999} (about which the Bloch vector {\boldmath $\rho$} precesses) initially lies along the $u$-axis. Upon applying a $3\pi/2$ pulse, the Bloch vector lies along the $-v$-axis, as indicated by {\boldmath $\rho'$} in Fig.~\ref{fig:rabiphase}(b). A $-\pi/2$ phase shift causes a $-\pi/2$ rotation of the torque vector, thus aligning the torque vector parallel to the Bloch vector. This corresponds to the entire population in the ground eigenstate $\phi_0$, which leaves the plane-wave populations fixed in time as we have demonstrated in our experiment. Alternatively, the combination of a $\pi/2$ pulse and $\pi/2$ phase shift also gives the same result. Anti-alignment of {\boldmath $\rho$} and {\boldmath $\Omega$} (for example a $\pi/2$ pulse followed by a $-\pi/2$ phase shift) would represent loading of the $\phi_1$ eigenstate. 

The apparent small modulation in the data after the application of the phase shift indicates that the eigenstate may not have been perfectly loaded. Such a modulation can arise due to slightly unequal populations in each momentum state at the time of the phase shift, and uncertainty in the magnitude of the phase shift. However, for our experiment the fluctuation in the data is within the shot-to-shot variation. This scheme for loading the ground state of a weak optical lattice can be realized for times as short as $t=\pi/(2\omega)$, which for our $4E_R$ deep lattice is 30~$\mu$s. This is much faster than the adiabatic method which can take on the order of milliseconds \cite{Orzel2001}.

\begin{figure}
  \subfigure[]{\includegraphics[width=0.6\linewidth]{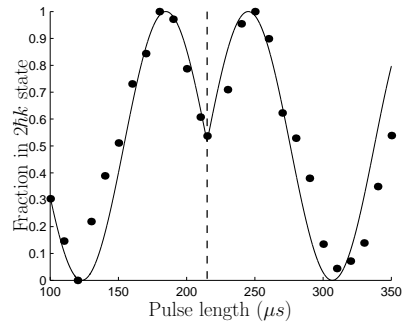}}  
  \subfigure[]{\includegraphics[width=0.6\linewidth]{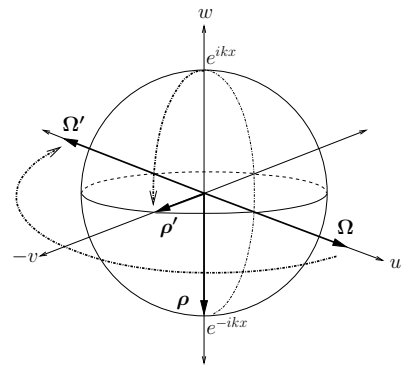}}
  \caption{\label{fig:rabiphase2} (a) The effect of a $-\pi$ phase shift to the optical lattice. The phase shift was applied at a time equivalent to a $3\pi$/2 pulse, indicated by the dashed line. (b) A representation of the experiment on a Bloch sphere.} 
\end{figure}

We now repeat the experiment shown in Fig.~\ref{fig:rabiphase} but with a $-\pi$ phase shift, and the result is shown in Fig.~\ref{fig:rabiphase2}(a). In this case, after the $-\pi$ phase shift the torque vector {\boldmath $\Omega'$} now lies along the $-u$-axis causing the Bloch vector to precess in the opposite direction, so that the cycling of population reverses. This is due to a $\pi$ phase change induced between $a_{0}$ and $a_{1}$, giving a wavefunction of the form
\begin{equation} \label{eq:psipi}
	\Psi(t\ge t_{\theta})  = \frac{1}{2}(1 - e^{i\omega(t-t_{\theta})})e^{-ikx} + \frac{1}{2}(1 + e^{i\omega(t-t_{\theta})})e^{ikx}.
\end{equation}

In summary, we have demonstrated non-adiabatic loading of a Bose-Einstein condensate into the ground state of a weak optical lattice at a quasi-momentum of $q = \hbar k$. Our method relies on applying a $\pm\pi/2$ phase shift to the lattice at a time when the populations in the two momentum components of the BEC are equal. Although this scheme does not inherently rely on the coherence of a condensate, its narrow momentum width ensures that the system is confined to a narrow region about the Brillouin zone boundary. We have used the cessation of Rabi cycling as the signature of having only one eigenstate populated. This phase-shifting technique can be used to manipulate the phase between the two Bloch bands of the optical lattice to give any superposition state, depending on the magnitude and timing of the phase shift. In the case of a weak optical lattice, the effect of such manipulations on Rabi cycling can be predicted using the Bloch sphere representation. 

\begin{acknowledgments}
The authors thank N.\ R.\ Thomas for assistance in the laboratory, K.\ J.\ Challis for help with computational work and R.\ J.\  Ballagh for useful comments. We acknowledge the support of the Marsden Fund of New Zealand, contract 02UOO080. C.\ M.\ acknowledges the support of the Foundation for Research Science and Technology, NZS{\&}T programme, contract UOOX0221.
\end{acknowledgments}


\begin{thebibliography}{31}
\expandafter\ifx\csname natexlab\endcsname\relax\def\natexlab#1{#1}\fi
\expandafter\ifx\csname bibnamefont\endcsname\relax
  \def\bibnamefont#1{#1}\fi
\expandafter\ifx\csname bibfnamefont\endcsname\relax
  \def\bibfnamefont#1{#1}\fi
\expandafter\ifx\csname citenamefont\endcsname\relax
  \def\citenamefont#1{#1}\fi
\expandafter\ifx\csname url\endcsname\relax
  \def\url#1{\texttt{#1}}\fi
\expandafter\ifx\csname urlprefix\endcsname\relax\def\urlprefix{URL }\fi
\providecommand{\bibinfo}[2]{#2}
\providecommand{\eprint}[2][]{\url{#2}}

\bibitem[{\citenamefont{Anderson et~al.}(1995)\citenamefont{Anderson, Ensher,
  Matthews, Wieman, and Cornell}}]{Anderson1995}
\bibinfo{author}{\bibfnamefont{M.~H.} \bibnamefont{Anderson}},
  \bibinfo{author}{\bibfnamefont{J.~R.} \bibnamefont{Ensher}},
  \bibinfo{author}{\bibfnamefont{M.~R.} \bibnamefont{Matthews}},
  \bibinfo{author}{\bibfnamefont{C.~E.} \bibnamefont{Wieman}},
  \bibnamefont{and} \bibinfo{author}{\bibfnamefont{E.~A.}
  \bibnamefont{Cornell}}, \bibinfo{journal}{Science}
  \textbf{\bibinfo{volume}{269}}, \bibinfo{pages}{198} (\bibinfo{year}{1995}).

\bibitem[{\citenamefont{Davis et~al.}(1995)\citenamefont{Davis, Mewes, Andrews,
  {van Druten}, Durfee, Kurn, and Ketterle}}]{Davis1995}
\bibinfo{author}{\bibfnamefont{K.~B.} \bibnamefont{Davis}},
  \bibinfo{author}{\bibfnamefont{M.-O.} \bibnamefont{Mewes}},
  \bibinfo{author}{\bibfnamefont{M.~R.} \bibnamefont{Andrews}},
  \bibinfo{author}{\bibfnamefont{N.~J.} \bibnamefont{{van Druten}}},
  \bibinfo{author}{\bibfnamefont{D.~S.} \bibnamefont{Durfee}},
  \bibinfo{author}{\bibfnamefont{D.~M.} \bibnamefont{Kurn}}, \bibnamefont{and}
  \bibinfo{author}{\bibfnamefont{W.}~\bibnamefont{Ketterle}},
  \bibinfo{journal}{Phys.\ Rev.\ Lett.} \textbf{\bibinfo{volume}{75}},
  \bibinfo{pages}{3969} (\bibinfo{year}{1995}).

\bibitem[{\citenamefont{Pethick and Smith}(2002)}]{Pethick2002}
\bibinfo{author}{\bibfnamefont{C.~J.} \bibnamefont{Pethick}} \bibnamefont{and}
  \bibinfo{author}{\bibfnamefont{H.}~\bibnamefont{Smith}},
  \emph{\bibinfo{title}{Bose-Einstein Condensation in Dilute Gases}}
  (\bibinfo{publisher}{Cambridge Univ. Press, Cambridge},
  \bibinfo{year}{2002}).

\bibitem[{\citenamefont{Anderson and Kasevich}(1998)}]{Anderson1998}
\bibinfo{author}{\bibfnamefont{B.~P.} \bibnamefont{Anderson}} \bibnamefont{and}
  \bibinfo{author}{\bibfnamefont{M.~A.} \bibnamefont{Kasevich}},
  \bibinfo{journal}{Science} \textbf{\bibinfo{volume}{282}},
  \bibinfo{pages}{1686} (\bibinfo{year}{1998}).

\bibitem[{\citenamefont{Kozuma et~al.}(1999)\citenamefont{Kozuma, Deng, Hagley,
  Wen, Lutwak, Helmerson, Rolston, and Phillips}}]{Kozuma1999}
\bibinfo{author}{\bibfnamefont{M.}~\bibnamefont{Kozuma}},
  \bibinfo{author}{\bibfnamefont{L.}~\bibnamefont{Deng}},
  \bibinfo{author}{\bibfnamefont{E.~W.} \bibnamefont{Hagley}},
  \bibinfo{author}{\bibfnamefont{J.}~\bibnamefont{Wen}},
  \bibinfo{author}{\bibfnamefont{R.}~\bibnamefont{Lutwak}},
  \bibinfo{author}{\bibfnamefont{K.}~\bibnamefont{Helmerson}},
  \bibinfo{author}{\bibfnamefont{S.~L.} \bibnamefont{Rolston}},
  \bibnamefont{and} \bibinfo{author}{\bibfnamefont{W.~D.}
  \bibnamefont{Phillips}}, \bibinfo{journal}{Phys.\ Rev.\ Lett.}
  \textbf{\bibinfo{volume}{82}}, \bibinfo{pages}{871} (\bibinfo{year}{1999}).

\bibitem[{\citenamefont{Stenger et~al.}(1999)\citenamefont{Stenger, Inouye,
  Chikkatur, Stamper-Kurn, Pritchard, and Ketterle}}]{Stenger1999}
\bibinfo{author}{\bibfnamefont{J.}~\bibnamefont{Stenger}},
  \bibinfo{author}{\bibfnamefont{S.}~\bibnamefont{Inouye}},
  \bibinfo{author}{\bibfnamefont{A.~P.} \bibnamefont{Chikkatur}},
  \bibinfo{author}{\bibfnamefont{D.~M.} \bibnamefont{Stamper-Kurn}},
  \bibinfo{author}{\bibfnamefont{D.~E.} \bibnamefont{Pritchard}},
  \bibnamefont{and} \bibinfo{author}{\bibfnamefont{W.}~\bibnamefont{Ketterle}},
  \bibinfo{journal}{Phys.\ Rev.\ Lett.} \textbf{\bibinfo{volume}{82}},
  \bibinfo{pages}{4569} (\bibinfo{year}{1999}).

\bibitem[{\citenamefont{Ovchinnikov et~al.}(1999)\citenamefont{Ovchinnikov,
  M{\"u}ller, Doery, Vredenbregt, Helmerson, Rolston, and
  Phillips}}]{Ovchinnikov1999}
\bibinfo{author}{\bibfnamefont{Y.~B.} \bibnamefont{Ovchinnikov}},
  \bibinfo{author}{\bibfnamefont{J.~H.} \bibnamefont{M{\"u}ller}},
  \bibinfo{author}{\bibfnamefont{M.~R.} \bibnamefont{Doery}},
  \bibinfo{author}{\bibfnamefont{E.~J.~D.} \bibnamefont{Vredenbregt}},
  \bibinfo{author}{\bibfnamefont{K.}~\bibnamefont{Helmerson}},
  \bibinfo{author}{\bibfnamefont{S.~L.} \bibnamefont{Rolston}},
  \bibnamefont{and} \bibinfo{author}{\bibfnamefont{W.~D.}
  \bibnamefont{Phillips}}, \bibinfo{journal}{Phys.\ Rev.\ Lett.}
  \textbf{\bibinfo{volume}{83}}, \bibinfo{pages}{284} (\bibinfo{year}{1999}).

\bibitem[{\citenamefont{Orzel et~al.}(2001)\citenamefont{Orzel, Tuchman,
  Fenselau, Yasuda, and Kasevich}}]{Orzel2001}
\bibinfo{author}{\bibfnamefont{C.}~\bibnamefont{Orzel}},
  \bibinfo{author}{\bibfnamefont{A.~K.} \bibnamefont{Tuchman}},
  \bibinfo{author}{\bibfnamefont{M.~L.} \bibnamefont{Fenselau}},
  \bibinfo{author}{\bibfnamefont{M.}~\bibnamefont{Yasuda}}, \bibnamefont{and}
  \bibinfo{author}{\bibfnamefont{M.~A.} \bibnamefont{Kasevich}},
  \bibinfo{journal}{Science} \textbf{\bibinfo{volume}{291}},
  \bibinfo{pages}{2386} (\bibinfo{year}{2001}).

\bibitem[{\citenamefont{Hensinger et~al.}(2001)\citenamefont{Hensinger,
  H{\"a}ffner, Browaeys, Heckenberg, Helmerson, McKenzie, Milburn, Phillips,
  Rolston, Rubinsztein-Dunlop et~al.}}]{Hensinger2001}
\bibinfo{author}{\bibfnamefont{W.~K.} \bibnamefont{Hensinger}},
  \bibinfo{author}{\bibfnamefont{H.}~\bibnamefont{H{\"a}ffner}},
  \bibinfo{author}{\bibfnamefont{A.}~\bibnamefont{Browaeys}},
  \bibinfo{author}{\bibfnamefont{N.~R.} \bibnamefont{Heckenberg}},
  \bibinfo{author}{\bibfnamefont{K.}~\bibnamefont{Helmerson}},
  \bibinfo{author}{\bibfnamefont{C.}~\bibnamefont{McKenzie}},
  \bibinfo{author}{\bibfnamefont{G.~J.} \bibnamefont{Milburn}},
  \bibinfo{author}{\bibfnamefont{W.~D.} \bibnamefont{Phillips}},
  \bibinfo{author}{\bibfnamefont{S.~L.} \bibnamefont{Rolston}},
  \bibinfo{author}{\bibfnamefont{H.}~\bibnamefont{Rubinsztein-Dunlop}},
  \bibnamefont{et~al.}, \bibinfo{journal}{Nature}
  \textbf{\bibinfo{volume}{412}}, \bibinfo{pages}{52} (\bibinfo{year}{2001}).

\bibitem[{\citenamefont{Cataliotti et~al.}(2001)\citenamefont{Cataliotti,
  Burger, Fort, Maddaloni, Minardi, Trombettoni, Smerzi, and
  Inguscio}}]{Cataliotti2001}
\bibinfo{author}{\bibfnamefont{F.~S.} \bibnamefont{Cataliotti}},
  \bibinfo{author}{\bibfnamefont{S.}~\bibnamefont{Burger}},
  \bibinfo{author}{\bibfnamefont{C.}~\bibnamefont{Fort}},
  \bibinfo{author}{\bibfnamefont{P.}~\bibnamefont{Maddaloni}},
  \bibinfo{author}{\bibfnamefont{F.}~\bibnamefont{Minardi}},
  \bibinfo{author}{\bibfnamefont{A.}~\bibnamefont{Trombettoni}},
  \bibinfo{author}{\bibfnamefont{A.}~\bibnamefont{Smerzi}}, \bibnamefont{and}
  \bibinfo{author}{\bibfnamefont{M.}~\bibnamefont{Inguscio}},
  \bibinfo{journal}{Science} \textbf{\bibinfo{volume}{293}},
  \bibinfo{pages}{843} (\bibinfo{year}{2001}).

\bibitem[{\citenamefont{Greiner et~al.}(2001)\citenamefont{Greiner, Bloch,
  Mandel, H{\"a}nsch, and Esslinger}}]{Greiner2001}
\bibinfo{author}{\bibfnamefont{M.}~\bibnamefont{Greiner}},
  \bibinfo{author}{\bibfnamefont{I.}~\bibnamefont{Bloch}},
  \bibinfo{author}{\bibfnamefont{O.}~\bibnamefont{Mandel}},
  \bibinfo{author}{\bibfnamefont{T.~W.} \bibnamefont{H{\"a}nsch}},
  \bibnamefont{and}
  \bibinfo{author}{\bibfnamefont{T.}~\bibnamefont{Esslinger}},
  \bibinfo{journal}{Appl.\ Phys.\ B} \textbf{\bibinfo{volume}{73}},
  \bibinfo{pages}{769} (\bibinfo{year}{2001}).

\bibitem[{\citenamefont{Cristiani et~al.}(2002)\citenamefont{Cristiani, Morsch,
  M{\"u}ller, Ciampini, and Arimondo}}]{Cristiani2002}
\bibinfo{author}{\bibfnamefont{M.}~\bibnamefont{Cristiani}},
  \bibinfo{author}{\bibfnamefont{O.}~\bibnamefont{Morsch}},
  \bibinfo{author}{\bibfnamefont{J.~H.} \bibnamefont{M{\"u}ller}},
  \bibinfo{author}{\bibfnamefont{D.}~\bibnamefont{Ciampini}}, \bibnamefont{and}
  \bibinfo{author}{\bibfnamefont{E.}~\bibnamefont{Arimondo}},
  \bibinfo{journal}{Phys.\ Rev.\ A} \textbf{\bibinfo{volume}{65}},
  \bibinfo{pages}{063612} (\bibinfo{year}{2002}).

\bibitem[{\citenamefont{Greiner et~al.}(2002)\citenamefont{Greiner, Mandel,
  Esslinger, H{\"a}nsch, and Bloch}}]{Greiner2002a}
\bibinfo{author}{\bibfnamefont{M.}~\bibnamefont{Greiner}},
  \bibinfo{author}{\bibfnamefont{O.}~\bibnamefont{Mandel}},
  \bibinfo{author}{\bibfnamefont{T.}~\bibnamefont{Esslinger}},
  \bibinfo{author}{\bibfnamefont{T.~W.} \bibnamefont{H{\"a}nsch}},
  \bibnamefont{and} \bibinfo{author}{\bibfnamefont{I.}~\bibnamefont{Bloch}},
  \bibinfo{journal}{Nature} \textbf{\bibinfo{volume}{415}}, \bibinfo{pages}{39}
  (\bibinfo{year}{2002}).

\bibitem[{\citenamefont{{Hecker Denschlag} et~al.}(2002)\citenamefont{{Hecker
  Denschlag}, Simsarian, H{\"a}ffner, McKenzie, Browaeys, Cho, Helmerson,
  Rolston, and Phillips}}]{HeckerDenschlag2002}
\bibinfo{author}{\bibfnamefont{J.}~\bibnamefont{{Hecker Denschlag}}},
  \bibinfo{author}{\bibfnamefont{J.~E.} \bibnamefont{Simsarian}},
  \bibinfo{author}{\bibfnamefont{H.}~\bibnamefont{H{\"a}ffner}},
  \bibinfo{author}{\bibfnamefont{C.}~\bibnamefont{McKenzie}},
  \bibinfo{author}{\bibfnamefont{A.}~\bibnamefont{Browaeys}},
  \bibinfo{author}{\bibfnamefont{D.}~\bibnamefont{Cho}},
  \bibinfo{author}{\bibfnamefont{K.}~\bibnamefont{Helmerson}},
  \bibinfo{author}{\bibfnamefont{S.~L.} \bibnamefont{Rolston}},
  \bibnamefont{and} \bibinfo{author}{\bibfnamefont{W.~D.}
  \bibnamefont{Phillips}}, \bibinfo{journal}{J.\ Phys.\ B:\ At.\ Mol.\ Opt.\
  Phys.} \textbf{\bibinfo{volume}{35}}, \bibinfo{pages}{3095}
  (\bibinfo{year}{2002}).

\bibitem[{\citenamefont{Jaksch et~al.}(1999)\citenamefont{Jaksch, Briegel,
  Cirac, Gardiner, and Zoller}}]{Jaksch1999}
\bibinfo{author}{\bibfnamefont{D.}~\bibnamefont{Jaksch}},
  \bibinfo{author}{\bibfnamefont{H.~J.} \bibnamefont{Briegel}},
  \bibinfo{author}{\bibfnamefont{J.~I.} \bibnamefont{Cirac}},
  \bibinfo{author}{\bibfnamefont{C.~W.} \bibnamefont{Gardiner}},
  \bibnamefont{and} \bibinfo{author}{\bibfnamefont{P.}~\bibnamefont{Zoller}},
  \bibinfo{journal}{Phys.\ Rev.\ Lett.} \textbf{\bibinfo{volume}{82}},
  \bibinfo{pages}{1975} (\bibinfo{year}{1999}).

\bibitem[{\citenamefont{Hemmerich}(1999)}]{Hemmerich1999}
\bibinfo{author}{\bibfnamefont{A.}~\bibnamefont{Hemmerich}},
  \bibinfo{journal}{Phys.\ Rev.\ A} \textbf{\bibinfo{volume}{60}},
  \bibinfo{pages}{943} (\bibinfo{year}{1999}).

\bibitem[{\citenamefont{Brennen et~al.}(1999)\citenamefont{Brennen, Caves,
  Jessen, and Deutsch}}]{Brennen1999}
\bibinfo{author}{\bibfnamefont{G.~K.} \bibnamefont{Brennen}},
  \bibinfo{author}{\bibfnamefont{C.~M.} \bibnamefont{Caves}},
  \bibinfo{author}{\bibfnamefont{P.~S.} \bibnamefont{Jessen}},
  \bibnamefont{and} \bibinfo{author}{\bibfnamefont{I.~H.}
  \bibnamefont{Deutsch}}, \bibinfo{journal}{Phys.\ Rev.\ Lett.}
  \textbf{\bibinfo{volume}{82}}, \bibinfo{pages}{1060} (\bibinfo{year}{1999}).

\bibitem[{\citenamefont{Burger et~al.}(2001)\citenamefont{Burger, Cataliotti,
  Fort, Minardi, Inguscio, Chiofalo, and Tosi}}]{Burger2001}
\bibinfo{author}{\bibfnamefont{S.}~\bibnamefont{Burger}},
  \bibinfo{author}{\bibfnamefont{F.~S.} \bibnamefont{Cataliotti}},
  \bibinfo{author}{\bibfnamefont{C.}~\bibnamefont{Fort}},
  \bibinfo{author}{\bibfnamefont{F.}~\bibnamefont{Minardi}},
  \bibinfo{author}{\bibfnamefont{M.}~\bibnamefont{Inguscio}},
  \bibinfo{author}{\bibfnamefont{M.~L.} \bibnamefont{Chiofalo}},
  \bibnamefont{and} \bibinfo{author}{\bibfnamefont{M.~P.} \bibnamefont{Tosi}},
  \bibinfo{journal}{Phys.\ Rev.\ Lett.} \textbf{\bibinfo{volume}{86}},
  \bibinfo{pages}{4447} (\bibinfo{year}{2001}).

\bibitem[{\citenamefont{Blakie and Ballagh}(2000)}]{Blakie2000}
\bibinfo{author}{\bibfnamefont{P.~B.} \bibnamefont{Blakie}} \bibnamefont{and}
  \bibinfo{author}{\bibfnamefont{R.~J.} \bibnamefont{Ballagh}},
  \bibinfo{journal}{J.\ Phys.\ B:\ At.\ Mol.\ Opt.\ Phys.}
  \textbf{\bibinfo{volume}{33}}, \bibinfo{pages}{3961} (\bibinfo{year}{2000}).

\bibitem[{\citenamefont{Allen and Eberly}(1975)}]{Allen1975}
\bibinfo{author}{\bibfnamefont{L.}~\bibnamefont{Allen}} \bibnamefont{and}
  \bibinfo{author}{\bibfnamefont{J.~H.} \bibnamefont{Eberly}},
  \emph{\bibinfo{title}{Optical Resonance and Two-Level Atoms}}
  (\bibinfo{publisher}{John Wiley and Sons, London}, \bibinfo{year}{1975}).

\bibitem[{\citenamefont{Hartmann and Hahn}(1962)}]{Hartmann1962}
\bibinfo{author}{\bibfnamefont{S.~R.} \bibnamefont{Hartmann}} \bibnamefont{and}
  \bibinfo{author}{\bibfnamefont{E.~L.} \bibnamefont{Hahn}},
  \bibinfo{journal}{Phys.\ Rev.} \textbf{\bibinfo{volume}{128}},
  \bibinfo{pages}{2042} (\bibinfo{year}{1962}).

\bibitem[{\citenamefont{Abragam}(1961)}]{Abragam1961}
\bibinfo{author}{\bibfnamefont{A.}~\bibnamefont{Abragam}},
  \emph{\bibinfo{title}{Principles of Nuclear Magnetism}}
  (\bibinfo{publisher}{Oxford Univ. Press, New York}, \bibinfo{year}{1961}).

\bibitem[{\citenamefont{Bai et~al.}(1985)\citenamefont{Bai, Yodh, and
  Mossberg}}]{Bai1985}
\bibinfo{author}{\bibfnamefont{Y.~S.} \bibnamefont{Bai}},
  \bibinfo{author}{\bibfnamefont{A.~G.} \bibnamefont{Yodh}}, \bibnamefont{and}
  \bibinfo{author}{\bibfnamefont{T.~W.} \bibnamefont{Mossberg}},
  \bibinfo{journal}{Phys.\ Rev.\ Lett.} \textbf{\bibinfo{volume}{55}},
  \bibinfo{pages}{1277} (\bibinfo{year}{1985}).

\bibitem[{\citenamefont{Cohen-Tannoudji and
  Reynard}(1977)}]{Cohen-Tannoudji1977}
\bibinfo{author}{\bibfnamefont{C.}~\bibnamefont{Cohen-Tannoudji}}
  \bibnamefont{and} \bibinfo{author}{\bibfnamefont{S.}~\bibnamefont{Reynard}},
  \bibinfo{journal}{J.\ Phys.\ B:\ At.\ Mol.\ Opt.\ Phys.}
  \textbf{\bibinfo{volume}{10}}, \bibinfo{pages}{345} (\bibinfo{year}{1977}).

\bibitem[{\citenamefont{Schmidt-Kaler et~al.}(2003)\citenamefont{Schmidt-Kaler,
  H{\"a}ffner, Riebe, Gulde, Lancaster, Deuschle, Becher, Roos, Eschner, and
  Blatt}}]{Schmidt-Kaler2003}
\bibinfo{author}{\bibfnamefont{F.}~\bibnamefont{Schmidt-Kaler}},
  \bibinfo{author}{\bibfnamefont{H.}~\bibnamefont{H{\"a}ffner}},
  \bibinfo{author}{\bibfnamefont{M.}~\bibnamefont{Riebe}},
  \bibinfo{author}{\bibfnamefont{S.}~\bibnamefont{Gulde}},
  \bibinfo{author}{\bibfnamefont{G.~P.~T.} \bibnamefont{Lancaster}},
  \bibinfo{author}{\bibfnamefont{T.}~\bibnamefont{Deuschle}},
  \bibinfo{author}{\bibfnamefont{C.}~\bibnamefont{Becher}},
  \bibinfo{author}{\bibfnamefont{C.~F.} \bibnamefont{Roos}},
  \bibinfo{author}{\bibfnamefont{J.}~\bibnamefont{Eschner}}, \bibnamefont{and}
  \bibinfo{author}{\bibfnamefont{R.}~\bibnamefont{Blatt}},
  \bibinfo{journal}{Nature} \textbf{\bibinfo{volume}{422}},
  \bibinfo{pages}{408} (\bibinfo{year}{2003}).

\bibitem[{\citenamefont{Cirac and Zoller}(1995)}]{Cirac1995}
\bibinfo{author}{\bibfnamefont{J.~I.} \bibnamefont{Cirac}} \bibnamefont{and}
  \bibinfo{author}{\bibfnamefont{P.}~\bibnamefont{Zoller}},
  \bibinfo{journal}{Phys.\ Rev.\ Lett.} \textbf{\bibinfo{volume}{74}},
  \bibinfo{pages}{4091} (\bibinfo{year}{1995}).

\bibitem[{\citenamefont{Metcalf and van~der Straten}(1999)}]{Metcalf1999}
\bibinfo{author}{\bibfnamefont{H.~J.} \bibnamefont{Metcalf}} \bibnamefont{and}
  \bibinfo{author}{\bibfnamefont{P.}~\bibnamefont{van~der Straten}},
  \emph{\bibinfo{title}{Laser Cooling and Trapping}}
  (\bibinfo{publisher}{Springer-Verlag, New York}, \bibinfo{year}{1999}).

\bibitem[{\citenamefont{Peik et~al.}(1997)\citenamefont{Peik, {Ben Dahan},
  Bouchoule, Castin, and Salomon}}]{Peik1997}
\bibinfo{author}{\bibfnamefont{E.}~\bibnamefont{Peik}},
  \bibinfo{author}{\bibfnamefont{M.}~\bibnamefont{{Ben Dahan}}},
  \bibinfo{author}{\bibfnamefont{I.}~\bibnamefont{Bouchoule}},
  \bibinfo{author}{\bibfnamefont{Y.}~\bibnamefont{Castin}}, \bibnamefont{and}
  \bibinfo{author}{\bibfnamefont{C.}~\bibnamefont{Salomon}},
  \bibinfo{journal}{Phys.\ Rev.\ A.} \textbf{\bibinfo{volume}{55}},
  \bibinfo{pages}{2989} (\bibinfo{year}{1997}).

\bibitem[{rap()}]{rapidload}
\bibinfo{note}{The initial degeneracy of the Bloch bands (when the lattice
  depth is zero) at $q=\hbar k$ ensures that an equal superposition is
  produced. This can be seen by projecting the initial wavefunction $e^{-ikx}$
  onto the Bloch states $\phi_0$ and $\phi_1$. An equal superposition occurs
  regardless of the lattice switch-on time, however it is convenient to rapidly
  switch on the lattice so that the phase between the eigenstates does not
  evolve significantly before the desired lattice depth is reached.}

\bibitem[{\citenamefont{Martin et~al.}(1999)\citenamefont{Martin, McKenzie,
  Thomas, Sharpe, Warrington, Manson, Sandle, and Wilson}}]{Martin1999}
\bibinfo{author}{\bibfnamefont{J.~L.} \bibnamefont{Martin}},
  \bibinfo{author}{\bibfnamefont{C.~R.} \bibnamefont{McKenzie}},
  \bibinfo{author}{\bibfnamefont{N.~R.} \bibnamefont{Thomas}},
  \bibinfo{author}{\bibfnamefont{J.~C.} \bibnamefont{Sharpe}},
  \bibinfo{author}{\bibfnamefont{D.~M.} \bibnamefont{Warrington}},
  \bibinfo{author}{\bibfnamefont{P.~J.} \bibnamefont{Manson}},
  \bibinfo{author}{\bibfnamefont{W.~J.} \bibnamefont{Sandle}},
  \bibnamefont{and} \bibinfo{author}{\bibfnamefont{A.~C.}
  \bibnamefont{Wilson}}, \bibinfo{journal}{J.\ Phys.\ B:\ At.\ Mol.\ Opt.\
  Phys.} \textbf{\bibinfo{volume}{32}}, \bibinfo{pages}{3065}
  (\bibinfo{year}{1999}).

\bibitem[{tra()}]{trapmod}
\bibinfo{note}{The tapered amplifier laser arrangement in \cite{Martin1999} has
  been replaced by an injection-locking scheme, and atoms are now continuously
  transferred in the double magneto-optical trap (MOT) arrangement.}

\end{thebibliography}
\end{document}